\documentstyle[12pt]{article}
\def\be{\begin{equation}}
\def\ee{\end{equation}}
\def\bea{\begin{eqnarray}}
\def\eea{\end{eqnarray}}

\begin{document}

\begin{center}
{\Large{\bf $(p+1)$-Algebra for Super $p$-Brane: 
the Nambu Bracket Reformulation\\}}

\vskip .5cm
{\large Davoud Kamani}
\vskip .1cm
{\it Faculty of Physics, Amirkabir University of Technology
(Tehran Polytechnic)\\
P.O.Box: 15875-4413, Tehran, Iran}\\
{\it e-mail: kamani@aut.ac.ir}\\
\end{center}

\begin{abstract}

We express the covariant actions of a super $p$-brane and the 
corresponding equations of motion, in the flat and curved 
superspaces, in terms of the Nambu $(p+1)$-brackets. These 
brackets make the $(p+1)$-algebra structure of super $p$-brane 
manifest. For the flat superspace, this reconstruction of 
the action also allows reformulating it in terms of two 
sets of differential forms.

\end{abstract}

\newpage
%%%%%%%%%%%%%%%%%%%%%%%%%%%%%%%%%%%%%%%%%%%%%%%%%%%%%%%%%%%%%%%%%%%%%%
\section{Introduction}

Recent studies reveal that M2-branes have a description in terms of a
3-algebra, a generalization of Lie algebra based on 
an antisymmetric triple product structure \cite{1}.
That is, 3-algebra relations have played an important role in the 
construction of the worldvolume theories of multiple M2-branes 
which have attracted considerable attention \cite{1,2}.
Various aspects of the 3-algebra can be seen 
in \cite{2} and the references therein.
However, the correspondence of the 2-algebra to the 
string theory and of the 3-algebra to the ${\cal{M}}$-theory 
can be understood from the dimensions of the string worldsheet and 
membrane worldvolume. This implies that the description of the 
super $p$-brane theory may require a $(p+1)$-algebra structure.
The cases of $p=1$ and $p=2$ have been worked out in \cite{3}.

This paper is dedicated to an important subject of construction 
of worldvolume theories for multiple $p$-branes. 
Recently this subject received a lot of attention due to the discovery 
of its relation to the multiple algebras. These algebras are defined in terms
of multiple commutators. The classical approximation to 
them is the well known Nambu multiple brackets.
So to explicitly formulate the brane action in terms of the 
multiple algebras the first step would be to rewrite the 
brane action in terms of the Nambu brackets. The Nambu $n$-brackets are 
a way for realizing the Lie $n$-algebra \cite{4}, which was developed 
by Filippov \cite{5}.

In Ref. \cite{6} it has been demonstrated that the (supersymmetric) 
$p$-brane action is invariant under the $(p + 1)$-dimensional diffeomorphisms.
In other words, there is an infinite-dimensional volume-preserving algebra 
of super $p$-branes. In this paper, we reformulate the super 
$p$-brane covariant action and the corresponding equations of motion, in the flat 
and curved superspaces, in terms of the 
Nambu $(p+1)$-brackets. Since the Nambu $(p+1)$-brackets are generators of the
$(p+1)$-dimensional diffeomorphisms, this reformulation reveals the above 
symmetry more explicitly. However, this reformulation represents the 
super $p$-branes on the basis of the $(p+1)$-algebra.

In fact, there are some advantages in reformulating of the membrane theory 
in terms of the 3-algebra. The same advantages also appear in  reformulating 
the $p$-brane theory in terms of the $(p+1)$-algebra. In addition, this 
reconstruction may provide a method for quantizing the theory. Beside, 
for the flat superspace, this reconstruction of the action enables us to
also reformulate it in terms of two sets of differential forms.  

This paper is organized as follows. In Sec. 2, we reconstruct a covariant,
$(p+1)$-algebra based action for a super $p$-brane in flat superspace.
In Sec. 3, we reformulate the super $p$-brane action 
in curved superspace in terms of the
$(p+1)$-algebra. In Sec. 4, quantizability of the theory will be discussed.
Section 5 is devoted to the conclusions.
%%%%%%%%%%%%%%%%%%%%%%%%%%%%%%%%%%%%%%%%%%%%%%%%%%%%%%%%%%%%%%%%%%%%%%%%%%
\section{The super $p$-brane in flat superspace,
on the basis of the $(p+1)$-algebra}

%%%%%%%%%%%%%%%%%%%%%%%%%%%%%%%%%%%%%%%%%%%%%%%%%%%%%%%%%%%%%%%%%%%%%%%%%%
\subsection{The action}

For the $(p+1)$-algebra description of 
a super $p$-brane propagating in the $D$-dimensional flat spacetime
we begin with the known action 
\bea
S_p=T_p\int d^{p+1} \sigma ({\cal{L}}_1 +{\cal{L}}_2),
\eea
\bea
&~& {\cal{L}}_1=\frac{1}{2(p+1)!}\phi^{-1}
\langle \Pi^{\mu_1},\Pi^{\mu_2},\cdot\cdot\cdot,\Pi^{\mu_{p+1}}\rangle
\langle \Pi_{\mu_1},\Pi_{\mu_2},\cdot\cdot\cdot,\Pi_{\mu_{p+1}}\rangle
-\frac{1}{2}\phi,
\nonumber\\
&~& {\cal{L}}_2=-\frac{2}{(p+1)!}\epsilon^{i_1\cdot\cdot\cdot i_{p+1}}
B_{i_1\cdot\cdot\cdot i_{p+1}}.
\eea 
The Lagrangian ${\cal{L}}_1$ is of the Schild type \cite{7}, i.e., 
to take off the square root of the Nambu-Goto action  
an auxiliary scalar field $\phi$ has been introduced.
${\cal{L}}_2$ is the Wess-Zumino Lagrangian. The degrees of 
freedom are: the spacetime coordinates $X^\mu$, the Majorana spinor
$\theta$ and the scalar field $\phi$.
The indices $\mu_1, \mu_2,\cdot\cdot\cdot ,\mu_{p+1}\in \{0,1,
\cdot\cdot\cdot ,D-1\}$
belong to the spacetime, while $i_1,i_2 \cdot\cdot\cdot,i_{p+1}\in
\{0,1,\cdot\cdot\cdot ,p\}$
indicate the $p+1$ directions of the brane worldvolume. 
The worldvolume coordinates are $\sigma^i$.
The Dirac matrices are denoted by $\Gamma^\mu$s. 
The metric of the spacetime is $\eta_{\mu\nu}={\rm diag}(-1,1, 
\cdot\cdot\cdot,1)$. The brane tension is given by the constant $T_p$.

The variable $\Pi^\mu_i$ has the definition 
\bea 
\Pi^\mu_i=\partial_i
X^\mu -i{\bar \theta}\Gamma^\mu\partial_i \theta,
\eea 
which is a supersymmetry invariant pull-back. In addition, we define 
\bea 
\langle\Pi^{\mu_1},\Pi^{\mu_2},\cdot\cdot\cdot,\Pi^{\mu_{p+1}}\rangle
=\epsilon^{i_1i_2\cdot\cdot\cdot i_{p+1}}
\Pi^{\mu_1}_{i_1}\Pi^{\mu_2}_{i_2}
\cdot\cdot\cdot\Pi^{\mu_{p+1}}_{i_{p+1}}, 
\eea
which is totally anti-symmetric.
%%%%%%%%%%%%%%%%%%%%%%%%%%%%%%%%%%%%%%%%%%%%%%%%%%%%%%%%%%%%%%%%%%%%%%%%%%%
\subsection{Equations of motion and symmetries}

The equations of motion have been extracted in \cite{6}. Since we want to
express them in terms of the Nambu brackets, we write them explicitly.
For the fields $\phi$, $X^\mu$ and $\theta$ the 
equations of motion are as in the following
\bea
&~& \phi-\sqrt{-g}=0,
\nonumber\\
&~& \partial_i (\sqrt{-g} g^{ij}\Pi^\mu_j)
-i\sqrt{-g}(-1)^{p(p+1)/2}\partial_i{\bar \theta} \Gamma^\mu 
\Gamma^{ij}\Gamma \partial_j \theta =0, 
\nonumber\\
&~& [1-(-1)^p \Gamma]\Gamma^i \partial_i \theta =0,
\eea
where the induced metric $g_{ij}$ is given by 
\bea 
g_{ij}=\Pi^\mu_i
\Pi^\nu_j \eta_{\mu\nu}. 
\eea 
The determinant of this metric is denoted by $g=\det g_{ij}$, which is  
\bea 
g=\frac{1}{(p+1)!}\langle
\Pi^{\mu_1}, \cdot\cdot\cdot,\Pi^{\mu_{p+1}} \rangle \langle
\Pi_{\mu_1}, \cdot\cdot\cdot, \Pi_{\mu_{p+1}}\rangle. 
\eea 
In addition, the matrices $\Gamma^i$, $\Gamma^{ij}$ and $\Gamma$ 
are defined as
\bea
&~& \Gamma^i=g^{ij}\Gamma_\mu \Pi^\mu_j ,
\nonumber\\
&~& \Gamma^{ij}=g^{ik}g^{jl}\Gamma_{\mu\nu} \Pi^\mu_k \Pi^\nu_l ,
\nonumber\\
&~& \Gamma=\frac{(-1)^{(p-2)(p-5)/4}}{(p+1)!\sqrt{-g}}
\Gamma_{\mu_1 \cdot\cdot\cdot
\mu_{p+1}} \langle \Pi^{\mu_1},\cdot\cdot\cdot ,
\Pi^{\mu_{p+1}}\rangle.
\eea
The matrix $\Gamma$ satisfies $\Gamma^2 =1$.
We shall see that the equations
of motion have expressions in terms of the $(p+1)$-algebra.

In addition to the worldvolume diffeomorphism invariance,
the action also is invariant under the following transformations
\bea
\delta \theta=\varepsilon, \;\;\;\;\;
\delta X^\mu =i {\bar \varepsilon}\Gamma^\mu \theta,
\;\;\;\;\;\delta \phi=0 ,
\eea
and
\bea
\delta_\kappa \theta=[1+(\phi/\sqrt{-g})\Gamma]\kappa (\sigma),\;\;\;\;\;
\delta_\kappa X^\mu=i{\bar \theta}\Gamma^\mu \delta_\kappa \theta,\;\;\;\;\;
\delta_\kappa \phi=4i\phi g^{ij}\Pi^\mu_i \partial_j {\bar \theta}
\Gamma_\mu \kappa(\sigma).
\eea 
The supersymmetry parameters $\varepsilon$ and $\kappa$ are 
spinors of the $D$-dimensional spacetime. The former is constant and the
later is local. 

By removing the auxiliary field $\phi$ through its equation 
of motion $\phi=\sqrt{-g}$, the Lagrangian ${\cal{L}}_1$ reduces to
the Nambu-Goto form
\bea
{\cal{L'}}_1=-\sqrt{-\det (\Pi^\mu_i \Pi^\nu_j \eta_{\mu\nu})} .
\eea
This Lagrangian also has a Polyakov expression
\bea
{\cal{L''}}_1=-\frac{1}{2}\sqrt{-h}[h^{ij}\Pi^\mu_i \Pi^\nu_j \eta_{\mu\nu}
-(p-1)],
\eea
where the {\it independent} auxiliary field $h_{ij}$ is the 
intrinsic worldvolume metric with $h=\det h_{ij}$. 
This is a convenient alternative  
form for ${\cal{L}}_1$. The equation of motion for $h_{ij}$, extracted from
(12), is 
\bea
h_{ij}=\Pi^\mu_i \Pi^\nu_j \eta_{\mu\nu}.
\nonumber
\eea
After eliminating $h_{ij}$ through its equation of motion, the 
Lagrangian (12) also reduces to
(11). Therefore, classically, ${\cal{L}}_1$, ${\cal{L'}}_1$ 
and ${\cal{L''}}_1$ are equivalent. 
However, ${\cal{L}}_2$ and the form (12) of ${\cal{L}}_1$ define
the Green-Schwarz action for the super $p$-brane. 
%%%%%%%%%%%%%%%%%%%%%%%%%%%%%%%%%%%%%%%%%%%%%%%%%%%%%%%%%%%%%%%%%%%%%%%%%%%
\subsection{The action based on the $(p+1)$-algebra}

The Nambu $(p+1)$-bracket of the variables $\phi_1, \cdot\cdot\cdot ,
\phi_{p+1}$ is defined by
\bea
\{\phi_1,\cdot\cdot\cdot ,\phi_{p+1}\}_{\rm N.B.}=
\epsilon^{i_1\cdot\cdot\cdot i_{p+1}}
\partial_{i_1}\phi_1 \cdot\cdot\cdot\partial_{i_{p+1}}\phi_{p+1}.
\eea
Therefore, in terms of the Nambu brackets the Eq. (4) takes the form
\bea
&~& \langle \Pi^{\mu_1},\Pi^{\mu_2},\cdot\cdot\cdot,\Pi^{\mu_{p+1}}\rangle
=\{X^{\mu_1},X^{\mu_2}, \cdot\cdot\cdot, X^{\mu_{p+1}}\}_{\rm N.B.}
\nonumber\\
&~& -i(p+1){\bar \theta}_\alpha\{(\Gamma^{[\mu_1}\theta)^\alpha,
X^{\mu_2}, \cdot\cdot\cdot, X^{\mu_{p+1}]}\}_{\rm N.B.}
\nonumber\\
&~& +\frac{p(p+1)}{2}{\bar \theta}_\alpha{\bar \theta}_\beta
\{(\Gamma^{[\mu_1}\theta)^\alpha,
(\Gamma^{\mu_2}\theta)^\beta,X^{\mu_3}, \cdot\cdot\cdot,
X^{\mu_{p+1}]}\}_{\rm N.B.} 
\nonumber\\
&~& -\frac{ip(p^2-1)}{6}{\bar \theta}_\alpha{\bar \theta}_\beta
{\bar \theta}_\gamma
\{(\Gamma^{[\mu_1}\theta)^\alpha,
(\Gamma^{\mu_2}\theta)^\beta,
(\Gamma^{\mu_3}\theta)^\gamma,X^{\mu_4}, \cdot\cdot\cdot,
X^{\mu_{p+1}]}\}_{\rm N.B.} 
\nonumber\\
&~& +\cdot\cdot\cdot+
\nonumber\\
&~& +i^{p+1} (-1)^{(p+1)(p+2)/2}{\bar \theta}_{\alpha_1}
{\bar \theta}_{\alpha_2}\cdot\cdot\cdot {\bar
\theta}_{\alpha_{p+1}}\{(\Gamma^{\mu_1} \theta)^{\alpha_1}, (\Gamma^{\mu_2}
\theta)^{\alpha_2},\cdot\cdot\cdot, (\Gamma^{\mu_{p+1}}
\theta)^{\alpha_{p+1}}\}_{\rm N.B.}
\nonumber\\
&~& =\sum^{p+1}_{n=0}\bigg{[}
\left(\begin{array}{c}
p+1 \\
n
\end{array}\right)
i^n (-1)^{n(n+1)/2}{\bar \theta}_{\alpha_1}
{\bar \theta}_{\alpha_2}\cdot\cdot\cdot {\bar\theta}_{\alpha_n}
\nonumber\\
&~& \times \{(\Gamma^{[\mu_1} \theta)^{\alpha_1}, (\Gamma^{\mu_2}
\theta)^{\alpha_2},\cdot\cdot\cdot, (\Gamma^{\mu_n}
\theta)^{\alpha_n},X^{\mu_{n+1}}, \cdot\cdot\cdot, 
X^{\mu_{p+1}]}\}_{\rm N.B.}\bigg{]}, 
\eea 
where the bracket $[\mu_1,\cdot\cdot\cdot, \mu_{p+1} ]$ indicates the 
anti-symmetrization of the indices.

Introducing Eq. (14) in the equations of motion (5) and the 
Lagrangian ${\cal{L}}_1$ we obtain the $(p+1)$-algebra expressions of them.
The explicit form of ${\cal{L}}_1$ is
\bea
&~& {\cal{L}}_1=\frac{1}{2(p+1)!} \phi^{-1}
\sum^{p+1}_{n=0}\sum^{p+1}_{m=0}\bigg{[}
\left(\begin{array}{c}
p+1 \\
n
\end{array}\right)
\left(\begin{array}{c}
p+1 \\
m
\end{array}\right)
\nonumber\\
&~& \times i^{m+n} (-1)^{(m+n)(m+n+1)/2}
{\bar \theta}_{\alpha_1}\cdot\cdot\cdot {\bar\theta}_{\alpha_n}
{\bar \theta}_{\beta_1}\cdot\cdot\cdot {\bar \theta}_{\beta_m}
\nonumber\\
&~& \times \{(\Gamma^{[\mu_1} \theta)^{\alpha_1}, 
\cdot\cdot\cdot, (\Gamma^{\mu_n}
\theta)^{\alpha_n},X^{\mu_{n+1}}, \cdot\cdot\cdot, 
X^{\mu_{p+1}]}\}_{\rm N.B.}
\nonumber\\
&~& \times \{(\Gamma_{[\mu_1} \theta)^{\beta_1}, 
\cdot\cdot\cdot, (\Gamma_{\mu_m}
\theta)^{\beta_m},X_{\mu_{m+1}}, \cdot\cdot\cdot, 
X_{\mu_{p+1}]}\}_{\rm N.B.}\bigg{]} -\frac{1}{2}\phi. 
\eea

In fact, the $B$-field can be expressed in terms of $X^\mu$s and 
$\theta$ as in the following \cite{8},
\bea
B_{i_1i_2\cdot\cdot\cdot i_{p+1}}=\frac{1}{2}\eta 
{\bar \theta} \Gamma_{\mu_1\cdot\cdot\cdot \mu_p}\partial_{i_{p+1}}\theta
\bigg{[} \sum^p_{r=0} i^{r+1}  
\left(\begin{array}{c}
p+1 \\
r+1
\end{array}\right)
({\bar \theta} \Gamma^{\mu_1}\partial_{i_1}\theta)\cdot\cdot\cdot
({\bar \theta} \Gamma^{\mu_r}\partial_{i_r}\theta)
\Pi^{\mu_{r+1}}_{i_{r+1}}\cdot\cdot\cdot\Pi^{\mu_p}_{i_p}
\bigg{]},
\eea
where $\eta$ is
\bea
\eta= (-1)^{(p-1)(p+6)/4}.
\nonumber
\eea
After eliminating the coefficients $B_{i_1i_2\cdot\cdot\cdot i_{p+1}}$ 
the Lagrangian ${\cal{L}}_2$ becomes 
\bea
&~& {\cal{L}}_2 =-\frac{\eta}{(p+1)!}\epsilon^{i_1\cdot\cdot\cdot i_{p+1}}
{\bar \theta} \Gamma_{\mu_1\cdot\cdot\cdot \mu_p}\partial_{i_{p+1}}\theta
\nonumber\\
&~& \times \bigg{[} \sum^p_{r=0} i^{r+1}  
\left(\begin{array}{c}
p+1 \\
r+1
\end{array}\right)
({\bar \theta} \Gamma^{\mu_1}\partial_{i_1}\theta)\cdot\cdot\cdot
({\bar \theta} \Gamma^{\mu_r}\partial_{i_r}\theta)
\Pi^{\mu_{r+1}}_{i_{r+1}}\cdot\cdot\cdot\Pi^{\mu_p}_{i_p}
\bigg{]}.
\eea
In a similar fashion to ${\cal{L}}_1$, the Lagrangian ${\cal{L}}_2$
in terms of the Nambu brackets has the expansion 
\bea
&~& {\cal{L}}_2=-\frac{1}{(p+1)!}\sum^p_{r=0}\sum^{p-r}_{m=0}\bigg{[}
\left(\begin{array}{c}
p+1 \\
r+1
\end{array}\right)
\left(\begin{array}{c}
p-r \\
m
\end{array}\right)
\nonumber\\
&~& \times i^{p-m+1} (-1)^{K_{r,m}}
{\bar \theta}_{\alpha_1} \cdot\cdot\cdot 
{\bar \theta}_{\alpha_r}{\bar \theta}_{\alpha_{r+m+1}}\cdot\cdot\cdot 
{\bar \theta}_{\alpha_p}{\bar \theta}_{\alpha_{p+1}}
\nonumber\\
&~& \times\{(\Gamma^{\mu_1}\theta)^{\alpha_1},\cdot\cdot\cdot,
(\Gamma^{\mu_r}\theta)^{\alpha_r}, X^{\mu_{r+1}},
\cdot\cdot\cdot, X^{\mu_{r+m}},
\nonumber\\
&~& (\Gamma^{\mu_{r+m+1}}\theta)^{\alpha_{r+m+1}},\cdot\cdot\cdot,
(\Gamma^{\mu_p}\theta)^{\alpha_p},
(\Gamma_{\mu_1\cdot\cdot\cdot \mu_p}\theta)^{\alpha_{p+1}}
\}_{\rm N.B.}\bigg{]}, 
\eea
where $K_{r,m}$ is 
\bea
K_{r,m}=p+\frac{1}{4}(p-1)(p+6)+\frac{1}{2}[r(r-1)+(p-r-m)(p+r-m+1)].
\eea

Let $Z^M =(X^\mu, \theta^\alpha)$ denote the coordinates of the
target space of the super $p$-brane. The worldvolume form 
$B_{i_1 \cdot\cdot\cdot i_{p+1}}$ is pull-back, i.e.,
\bea
B_{i_1 \cdot\cdot\cdot i_{p+1}}=\partial_{i_1}Z^{M_1}\cdot\cdot\cdot
\partial_{i_{p+1}}Z^{M_{p+1}}B_{M_{p+1}\cdot\cdot\cdot M_{1}},
\eea
where $B_{M_{p+1}\cdot\cdot\cdot M_{1}}$ are components of a 
$(p+1)$-form potential in the superspace. Therefore, an other
$(p+1)$-algebra expression of ${\cal{L}}_2$ is
\bea
{\cal{L}}_2=-\frac{2}{(p+1)!}\{ Z^{M_{1}}, \cdot\cdot\cdot, 
Z^{M_{p+1}}\}_{\rm N.B.}B_{M_{p+1}\cdot\cdot\cdot M_{1}}.
\eea

According to the Eqs. (15), (18) and (21), 
all derivatives have been absorbed in the Nambu $(p+1)$-brackets. 
Hence, the $(p+1)$-algebra structure is made manifest.
%%%%%%%%%%%%%%%%%%%%%%%%%%%%%%%%%%%%%%%%%%%%%%%%%%%%%%%%%%%%%%%%%%%%%%%%%%%%
\subsection{The action in terms of differential forms}

The $(p+1)$-algebra description of a super $p$-brane enables us to write
the action $S_p=S^{(1)}_p+S^{(2)}_p$ in the language of differential forms
\bea
&~& S^{(1)}_p=\frac{T_p}{2} \sum^{p+1}_{n=0}\sum^{p+1}_{m=0}\bigg{\{}
\left(\begin{array}{c}
p+1 \\
n
\end{array}\right)
\left(\begin{array}{c}
p+1 \\
m
\end{array}\right)
i^{m+n} (-1)^{(m+n)(m+n+1)/2}\int_{\rm w.v.} \phi^{-1} A_{(m,n)}\bigg{\}}
\nonumber\\
&~& \;\;\;\;\;\;\;\;\;
-\frac{T_p}{2} \int_{\rm w.v.} d^{p+1} \sigma \phi ,
\nonumber\\
&~& S^{(2)}_p=-T_p \sum^{p}_{r=0}\sum^{p-r}_{m=0}\bigg{\{}
\left(\begin{array}{c}
p+1 \\
r+1
\end{array}\right)
\left(\begin{array}{c}
p-r \\
m
\end{array}\right)
i^{p-m+1} (-1)^{K_{r,m}}\int_{\rm w.v.} C_{(r,m)}\bigg{\}},
\eea
The differential $(p+1)$-forms are defined by
\bea
&~& A_{(m,n)}=
\frac{1}{(p+1)!}\{ Y_{[\mu_1}, \cdot \cdot \cdot , Y_{\mu_m},
X_{\mu_{m+1}}, \cdot \cdot \cdot , X_{\mu_{p+1}]} \}_{\rm N.B.} \times
\nonumber\\
&~& \;\;\;\;\;\;\;\;\;\;\;\;\;\;(dY^{\mu_1} 
\wedge \cdot \cdot \cdot \wedge dY^{\mu_n}\wedge dX^{\mu_{n+1}}
\wedge \cdot \cdot \cdot \wedge dX^{\mu_{p+1}})|_{\rm w.v.},
\nonumber\\
&~& C_{(r,m)}=\frac{1}{(p+1)!}
(dY^{\mu_1} \wedge \cdot \cdot \cdot \wedge dY^{\mu_r}\wedge dX^{\mu_{r+1}}
\wedge \cdot \cdot \cdot \wedge dX^{\mu_{r+m}}
\nonumber\\
&~& \;\;\;\;\;\;\;\;\;\;\;\;\;\;\wedge 
dY^{\mu_{r+m+1}} \wedge \cdot \cdot \cdot 
\wedge dY^{\mu_p}\wedge d Z_{\mu_1 \cdot \cdot \cdot \mu_p})|_{\rm w.v.}.
\eea
where the restriction $|_{\rm w.v.}$ means pull-back of wedge products 
on the worldvolume of the super $p$-brane, e.g. 
\bea
dX^\mu|_{\rm w.v.}=\partial_i X^\mu d \sigma^i.
\nonumber
\eea
The variable $Y^\mu$ and the antisymmetric tensor 
$Z_{\mu_1 \cdot \cdot \cdot \mu_p}$ are given by
\bea
&~& Y^\mu={\bar \theta} \Gamma^\mu \theta ,
\nonumber\\
&~& Z_{\mu_1 \cdot \cdot \cdot \mu_p}={\bar \theta} 
\Gamma_{\mu_1\cdot \cdot \cdot \mu_p} \theta .
\eea
These wedge products define differential $(p+1)$-forms.
The components of these forms explicitly have been given in terms
of the coordinates $\{X^\mu\}\bigcup \{\theta^\alpha\}$. These forms 
do not have the pure bosonic part, i.e., they vanish in the absence of 
$\theta^\alpha$s. Hence, they exist only for the super branes.

Since $\{X^\mu (\tau; \sigma^1, \cdot \cdot \cdot , \sigma^p) \} \bigcup 
\{\theta^\alpha (\tau; \sigma^1, \cdot \cdot \cdot , \sigma^p) \}$ 
are coordinates of the worldvolume 
of the super $p$-brane in the superspace, the actions
$S^{(1)}_p$ and $S^{(2)}_p$ imply that the super $p$-brane is coupled to
the potential forms 
\bea
&~&\{A_{(m,n)}|m,n=0,1,\cdot \cdot \cdot,p+1 \},
\nonumber\\ 
&~&\{C_{(r,m)}|m=0,1, \cdot \cdot \cdot, p-r ;\;r=0,1,\cdot \cdot \cdot,p\}.
\nonumber
\eea
In fact, only reformulating the super $p$-brane action on the basis of the 
$(p+1)$-algebra reveals these differential forms.
%%%%%%%%%%%%%%%%%%%%%%%%%%%%%%%%%%%%%%%%%%%%%%%%%%%%%%%%%%%%%%%%%%%%%%%%%%%%
\section{The super $p$-brane in the curved superspace}

We assume the target space of the super $p$-brane to be a
curved supermanifold with $E^A_M (Z)$ as its corresponding supervielbeins.
The $A=a,\;\alpha$ are the tangent-space indices.
Then the super $p$-brane action is given by 
\bea
I_p=-T_p \int d^{p+1}\sigma \bigg{(}\sqrt{-\det(E^a_iE^b_j \eta_{ab})} 
+\frac{2}{(p+1)!} \epsilon^{i_1
\cdot\cdot\cdot i_{p+1}} E^{A_1}_{i_1}\cdot\cdot\cdot
E^{A_{p+1}}_{i_{p+1}} B_{A_{p+1}\cdot\cdot\cdot A_1}\bigg{)},
\eea
where 
\bea
E^A_i=\partial_i Z^M E^A_M, 
\eea
is the pull-back of the supervielbeins $E^A_M$. 
The field $B_{A_{p+1}\cdot\cdot\cdot A_1}(Z)$ is a
superspace $(p+1)$-form potential. Note that due to the $\kappa$-symmetry 
of the action, only special values of $p$ and $D$ are allowed, (see  
Ref. \cite{9} and the references therein).

In this action, the $(p+1)$-algebra can also be introduced.
Since we have
\bea
&~& \det (E^a_i E^b_j \eta_{ab})=\frac{1}{(p+1)!}
\langle E^{a_1}, \cdot\cdot\cdot , E^{a_{p+1}}\rangle
\langle E_{a_1}, \cdot\cdot\cdot,E_{a_{p+1}}\rangle,
\nonumber\\
&~& \langle E^{a_1}, \cdot\cdot\cdot ,E^{a_{p+1}}\rangle
=\epsilon^{i_1 \cdot\cdot\cdot i_{p+1}}
E^{a_1}_{i_1}\cdot\cdot\cdot E^{a_{p+1}}_{i_{p+1}},
\eea
the action (22) can be reformulated in terms of the Nambu $(p+1)$-brackets
\bea
&~& I_p =-T_p\int d^{p+1}\sigma\bigg{\{}\bigg{(}-\frac{1}{(p+1)!}
E^{a_1}_{M_1}\cdot\cdot\cdot E^{a_{p+1}}_{M_{p+1}}
E^{b_1}_{N_1}\cdot\cdot\cdot E^{b_{p+1}}_{N_{p+1}}
\nonumber\\
&~& \times \{Z^{M_1},\cdot\cdot\cdot, Z^{M_{p+1}}\}_{\rm N.B.}
\{Z^{N_1},\cdot\cdot\cdot, Z^{N_{p+1}}\}_{\rm N.B.}
\eta_{a_1b_1}\cdot\cdot\cdot \eta_{a_{p+1}b_{p+1}}\bigg{)}^{1/2}
\nonumber\\
&~& +\frac{2}{(p+1)!}
E^{A_1}_{M_1}\cdot\cdot\cdot E^{A_{p+1}}_{M_{p+1}}
\{ Z^{M_1},\cdot\cdot\cdot, Z^{M_{p+1}} \}_{\rm N.B.}
B_{A_{p+1}\cdot\cdot\cdot A_1}\bigg{\}}.
\eea
The novelty of this reformulation is the appearance of the $(p+1)$-algebra.
%%%%%%%%%%%%%%%%%%%%%%%%%%%%%%%%%%%%%%%%%%%%%%%%%%%%%%%%%%%%%%%%%%%%%%%%%%%%
\section{A note on the quantization of the reformulated actions}

Due to the intrinsic nonlinearities,
quantization of $p$-branes is a difficult problem. 
There are different quantum mechanical approaches associated 
with the quantum dynamics of $p$-branes based on viewing $p$-branes 
as gauge theories of volume-preserving diffeomorphisms.
In other words, several quantum mechanical methods for $p$-branes 
are proposed based on the role that the volume-preserving diffeomorphisms 
group has on the physics of these extended objects. 

The other experience is the quantum Nambu brackets. They describe 
the quantum behavior of systems equivalently to the standard Hamiltonian 
quantization. For example, they serve to guide quantization of more general 
even-dimensional topological branes \cite{10}.

Thus, by appropriate replacing of the classical Nambu brackets with
the quantum Nambu brackets, one may achieve the quantization of the 
reformulated actions in this paper. This is not straightforward. 
It seems the statue of quantizability of the reformulated actions is 
that they are not quantizable for $p \neq 1$, for the same reason that 
a quantum membrane theory has yet to be formulated.
%%%%%%%%%%%%%%%%%%%%%%%%%%%%%%%%%%%%%%%%%%%%%%%%%%%%%%%%%%%%%%%%%%%%%%%%%%%
\section{Conclusions}

In the first part of this paper, we expressed the super $p$-brane 
action and the corresponding equations of
motion, in the flat superspace, in terms of the Nambu $(p+1)$-brackets.
In the second part, for a super $p$-brane which lives in a 
curved superspace, we obtained the Nambu $(p+1)$-bracket expression of the
action. This reformulation is another language for describing the
super $p$-branes and gives a new insight on the branes. It may provide
a way for quantizing the $p$-branes. 

In both the above cases, all derivatives appeared through the Nambu 
$(p+1)$-brackets and hence the $(p+1)$-algebra structure 
for the super $p$-brane theory was made manifest. 
This is related to the fact that: 1) the
(supersymmetric) $p$-brane action is invariant under the $(p+1)$-dimensional
diffeomorphisms and 2) the Nambu $(p+1)$-brackets are generators of the
$(p+1)$-dimensional diffeomorphisms.

Finally, for flat superspace, we found two sets of differential 
$(p+1)$-forms that couple to the super $p$-brane. This result originates from  
the reformulation and cannot be seen in the original form of the action.
%%%%%%%%%%%%%%%%%%%%%%%%%%%%%%%%%%%%%%%%%%%%%%%%%%%%%%%%%%%%%%%%%%%%%%%%%%%%

\end{document}